\documentclass[apjl]{emulateapj}

\usepackage{natbib}
\usepackage{enumerate}
\usepackage{color}
\newcommand{\La}{Lyman-$\alpha$}

\begin{document}

\title{Transition disk chemistry and future prospects with ALMA}
    
    \author{L. Ilsedore Cleeves, Edwin A. Bergin, Thomas J. Bethell, Nuria Calvet, and Jeffrey K. J. Fogel}
    \affil{Department of Astronomy, University of Michigan, 825 Dennison Building, 500 Church St, Ann Arbor, MI 48109}
    
    \and
    
   \author{J\"{u}rgen Sauter and Sebastian Wolf}
    \affil{Christian-Albrechts-Universit\"{a}t zu Kiel, Institut f\"{u}r Theoretische Physik und Astrophysik, Leibnizstr. 15, 24098 Kiel, Germany}

\begin{abstract}

We explore the chemical structure of a disk that contains a large central gap of R $\sim$ 45 AU, as is commonly seen in transitional disk systems.  
In our chemical model of a disk with a cleared inner void, the midplane
becomes revealed to the central star so that it is directly irradiated.  The midplane material at the truncation radius is permissive to reprocessed optical heating radiation, but opaque to the photo-dissociating ultraviolet, creating an environment abundant in gas-phase molecules.  Thus the disk midplane, which would otherwise for a full disk be dominated by near complete heavy element freeze-out, should become observable in molecular emission.  If this prediction is correct this has exciting prospects for observations with the Atacama Large Millimeter/Submillimeter Array (ALMA), as the inner transition region should thus be readily detected and resolved, especially using high-J rotational transitions excited in the high density midplane gas. Therefore such observations will potentially provide us with a direct probe of the physics and chemistry at this actively evolving interface.

\end{abstract}


\section{Introduction}

Over the last thirty years, our knowledge of pre-main sequence evolution has undergone significant advances.  Observations of the full spectral energy distribution (SED) from disks around young stars have shown that not all are alike, with the striking discovery of a subset of disks with optically thin inner ``holes'' devoid of small grains, surrounded by optically thick outer disks \citep[e.g.][]{strom1989,calvet2005,cieza2010,espaillat2010}.  This inner void was later confirmed by resolved sub-mm interferometry \citep[e.g.][]{pietu2006,Brown2008,hughes2009, andrews2009, andrews2011} and was interpreted as an intermediate stage between primordial and debris disks, coined ``transition disks'' \citep{strom1989}.  

In this work we adopt the physical definition of ``transition disk'' as an optically thick disk truncated within some inner radius within which there has either been substantial grain growth or removal.  Such objects are of particular interest, as the presence of a gap has been attributed to clearing by young protoplanets \citep[e.g.][]{skrutskie1990,bryden1999,rice2003} or tidal interactions with young stellar companions \citep[e.g.][]{dutrey1994}.  

\citet{dullemond2001} investigated the physical structure of Herbig Ae/Be disks that possess large inner radii similar to those seen in low mass transition disks. In this work it was found that at the inner rim, the normal incidence angle of the stellar irradiation causes the disk rim to be much hotter than it would otherwise be for a classical flared disk, where the radiation arrives at a glancing angle. Further observational evidence for the directly irradiated wall was later seen both in studies of the SED \citep[e.g. ][]{espaillat2007} as well as through scattered light imaging \citep{Brown2008}.  

ALMA will readily resolve such inner gaps and voids in the dust disk \citep[e.g. ][]{wolf2002,wolf2005}; however, here we seek to make predictions regarding the significant gas reservoir in transition disks. In particular, one interesting aspect of disk chemistry that has yet to be explored is the potential that this warm UV irradiated inner rim should have unique chemical properties.  This has important implications since it is possible that the physical and chemical conditions of the previously hidden, but now exposed, midplane will be revealed and potentially detectable with high spatial resolution observations, such as those anticipated by ALMA.  Here we present a chemical model of a protoplanetary disk with a large inner gap similar to those seen in classical transition disks \citep[e.g.][]{hughes2009,espaillat2010} as a prospective study of observability with ALMA.

\section{Chemical Modeling} \label{sec:intro}

\subsection{Disk Framework} \label{sec:mod}

For our model we adopt the comprehensive disk structure from \citet{sauter2009}, originally purposed for CB26 and constrained by a large set of observations:  the SED from nanometer to millimeter wavelengths, near-infrared scattered light images, and resolved millimeter images.  For the central star, the model of \citet{sauter2009} assumes standard T Tauri values of M $=0.5$ M$_\odot$ and L $=0.9$ L$_\odot$.  The disk density profile is given by:

\begin{equation}
\rho=\rho_{0}{\Bigg(\frac{r}{r_{0}}\Bigg)}^{-\alpha}\exp\Bigg(-\frac{1}{2}\Bigg(\frac{z}{h}\Bigg)^2\Bigg)
\end{equation}
\begin{equation}
h=h_{0}{\Bigg(\frac{r}{r_{0}}\Bigg)}^{\beta} 
\end{equation}
Using this model and temperatures derived with MC3D \citep{wolf1999}, \citet{sauter2009} successfully reproduced the full set of observations with the following best-fit model parameters: $r_0=100$ AU, $h_{0}=10$ AU, $\alpha=2.2$, $\beta=1.4$, $R_{\rm outer}=200$ AU and $R_{\rm inner}=45$ AU.  The measured size of the hole is typical of that seen in resolved millimeter observations and SED modeling of transition disks, which range from a few AU to upwards of 70 AU \citep{andrews2011,espaillat2007}. The inner disk irradiation substantially increases the disk temperature at the wall, reaching $\sim50$ K, as compared to a typical midplane temperature of $\sim15$ K as seen at larger radii.  We note that this model is taken as a ``snapshot'' and any subsequent disk physical evolution is beyond the scope of this work, though the chemistry of a variety of disk geometries will be explored in an upcoming paper.

\begin{figure}
\epsscale{1.2}
\plotone{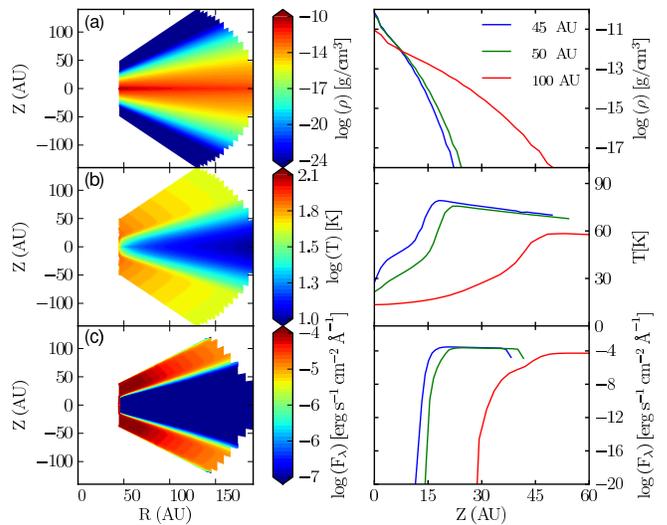}
\caption{{\bf Left:}  Plot of the transition disk model with the central star at the origin.  Panels:  a) gas density b) temperature c) FUV radiation field at $1600$ \AA. {\bf Right:} Vertical cuts of the same quantities taken at R = 45, 50 and 100 AU. \label{fig1}}
\end{figure}

For the gas density we assume a standard ISM gas-to-dust mass ratio of $f_{\rm g}=100$ and that the dust and gas are co-spatial.  The gas temperature is taken to be equal to T$_{\rm dust}$, which holds true near the dense midplane \citep{jonkheid2004}.  We furthermore note T$_{\rm dust}$ was originally derived assuming passive heating by the central star, with accretion heating treated as negligible \citep{sauter2009}.  The effects of inclusion of accretion heating are discussed in Section~\ref{sec:accre}. The model physical structure is shown in Fig. \ref{fig1}.  

For the opacities we adopt \citet{WD2001} for a blend of astronomical silicates and carbonaceous grains with R$_{\rm V}=5.5$.  We also assume an unsettled disk and uniform dust composition for simplicity.  Deviations from these assumptions, such as the affect of vertical settling of small grains and grain-growth, are discussed in Section~\ref{sec:dustgr}.  

\subsection{Radiation Field} \label{sec:rad}
Both the ultraviolet and X-ray radiation field are believed to be dominant factors driving the chemistry in disks \citep{glassgold1997,glassgold2004,vanzadelhoff2003}.  While low mass young stars peak in the optical regime with negligible chromospheric contribution to the FUV, the accretion shock at the central star provides a significant source of both FUV and soft X-ray flux \citep[e.g. ][]{gunther2007}.  We neglect external contribution from the ISRF since the stellar FUV will dominate external sources by orders of magnitude in the inner disk \citep{bergin2003}, though its import may lie in understanding the outer disk chemistry \citep{oberg2010}.   Due to the geometry of the star-disk system it is then necessary to properly treat the radiative transfer into the disk \citep{willacyandlanger2000,vanzadelhoff2003,fogel2011}.  This is especially true in transition disks, where voids and gaps can allow photons to propagate more freely into the outer disk material.  Possible implications of the presence of a small amount of undetected dust within the gap are discussed in Section~\ref{sec:innerdust}.

\subsubsection{Continuum Radiative Transfer}  \label{sec:contrad}

We assume the FUV continuum opacity is dominated by the dust, and consequently the UV field is dependent on settling of small grains, opacities assumed, and disk geometry.   Using the measured FUV spectrum of TW Hydra \citep{herczeg2002,Herczeg2004} binned down to nine discrete wavelengths between 950-2000 \AA, we calculate the continuum radiative transfer into the disk using the method of \citet{bethell2011}, implementing the opacities described in Section~\ref{sec:mod}.  The choice of nine wavelengths was motivated by the need to sufficiently capture the shape of the FUV continuum, to be insensitive to individual weak emission lines, and to also remain computationally efficient.    

\subsubsection{\La\ Radiative Transfer} \label{sec:larad}

While many individual lines in the observed TW Hydra spectrum are weak, the \La\ line alone carries $\sim85\%$ of the total FUV flux \citep{Herczeg2004, bergin2003}. This line is furthermore expected to play a significant chemical role as a number of molecular species have photodissociation cross sections near 1216\AA\ \citep{fogel2011}.  In addition to dust scattering, \La\ undergoes isotropic scattering off hydrogen atoms \citep{bethell2011}, which in principle requires an iterative and computationally expensive calculation between the radiation field and the chemistry.  

\begin{figure}
\epsscale{1.17}
\plotone{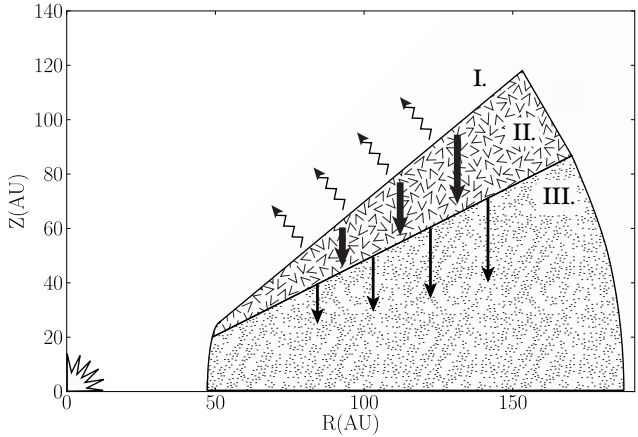}
\caption{Ly$\alpha$ schematic illustration.  (I.) ``Free-streaming'' region above $\tau_{\rm radial}\sim1$ surface;  half of the Ly$\alpha$ radiation escapes the disk.  (II.) H-scattering layer with a vertical depth of $\tau_{\rm vertical}\sim1$.  Remaining photons diffuse downward until they escape the atomic region. (III.) Region where H is predominantly H$_2$.  Dust now becomes the dominant source of opacity to Ly$\alpha$ photons, and the Ly$\alpha$ radiation now behaves as vertically attenuated continuum photons, which significantly enhances its penetrating power.  The (I.) $\rightarrow$ (II.) interface illustrates the $\tau \sim1$ surface for continuum photons, whereas the (II.) $\rightarrow$ (III.) interface is effectively the $\tau \sim1$ surface for Ly$\alpha$ photons.  \label{fig3}}
\end{figure}

In this work, we incorporate the effects of \La\ using an approximate treatment motivated by \citet{bethell2011}.  The procedure is illustrated in Fig. \ref{fig3}.  If one assumes a priori that there is an optically thick layer of atomic hydrogen on the disk surface located at $\tau_{\rm FUV}=1$ as defined by the dust (Fig. \ref{fig3}: Region II), the \La\ photons will first encounter this isotropic H-scattering layer.  
A fraction of the radiation will be lost to space ($\lesssim50\%$, Region I:  wavy arrows), with the surviving \La\ photons diffusively propagating through the H-layer.  Below the atomic layer, the hydrogen is predominantly H$_2$ and the \La\ photons proceed as continuum photons scattering off only the dust grains (Fig. \ref{fig3}: Region III).  Thus, at the base of the hydrogen scattering layer (II$\rightarrow$III), the photons effectively form a layer of isotropically emitting point sources, which together create a planar source of radiation that shines at a nearly normal angle to the midplane.  Therefore even though a significant fraction of \La\ photons are lost to space, the remaining photons will have greater vertical penetration power into the disk and can propagate many AU deeper than the UV continuum radiation.  On the front edge of the disk, at the dense inner rim, \La\ is similarly ``stopped'' as was seen for the continuum photons. (cf. Fig. \ref{fig1} (c)).   We discuss the implications of an inner void that is not empty of gas on \La\ transfer in Section~\ref{sec:innergas}. 

\subsection{Reaction Network} \label{sec:react}
Combining the model detailed in Section~\ref{sec:mod} and the radiation field in Section~\ref{sec:rad}, the resultant chemistry is calculated using \citet{fogel2011}'s comprehensive disk chemical model, based on the Ohio State University Astrophysical Chemistry Group's gas-phase network \citep{smith2004}. The reaction types include photo-desorption, photo-dissociation, freeze-out, grain surface reactions, ion and electron reactions, cosmic-ray and stellar X-ray ionization, and radiative reactions.  In total, the network encompasses 5910 reactions and 639 reacting species, including some time-dependent reactions, encompassing the main species of astrochemical importance, and described in detail in \citet{fogel2011}.  

The model initially assumes uniform molecular cloud chemical abundances \citet{aikawa1999} and follows the chemical evolution for 3 Myr.  In this work we take the abundances at 1 Myr, which is long enough such that the chemistry has ``relaxed'' but not so long that the disk would likely have {\em physically} evolved away from this state.  

Furthermore, we assume a typical integrated T Tauri star X-ray luminosity of 10$^{30}$ erg s$^{-1}$ and a thermal X-ray spectrum between 1-10 keV \citep[][and references therein]{glassgold1997}.  The model of \citet{fogel2011} incorporates the method of \citet{aikawa2001} for X-ray propagation.  For cosmic rays we adopt a typical cosmic ray ionization rate of 1.3 $\times$ 10$^{-17}$ s$^{-1}$ per H, with an attenuation column of 96 g cm$^{-2}$ \citep{umebayashi1981}.  

\section{Results} \label{sec:result}

\subsection{Disk Chemistry} \label{sec:diskchem}
\begin{figure*}
\epsscale{1}
\plotone{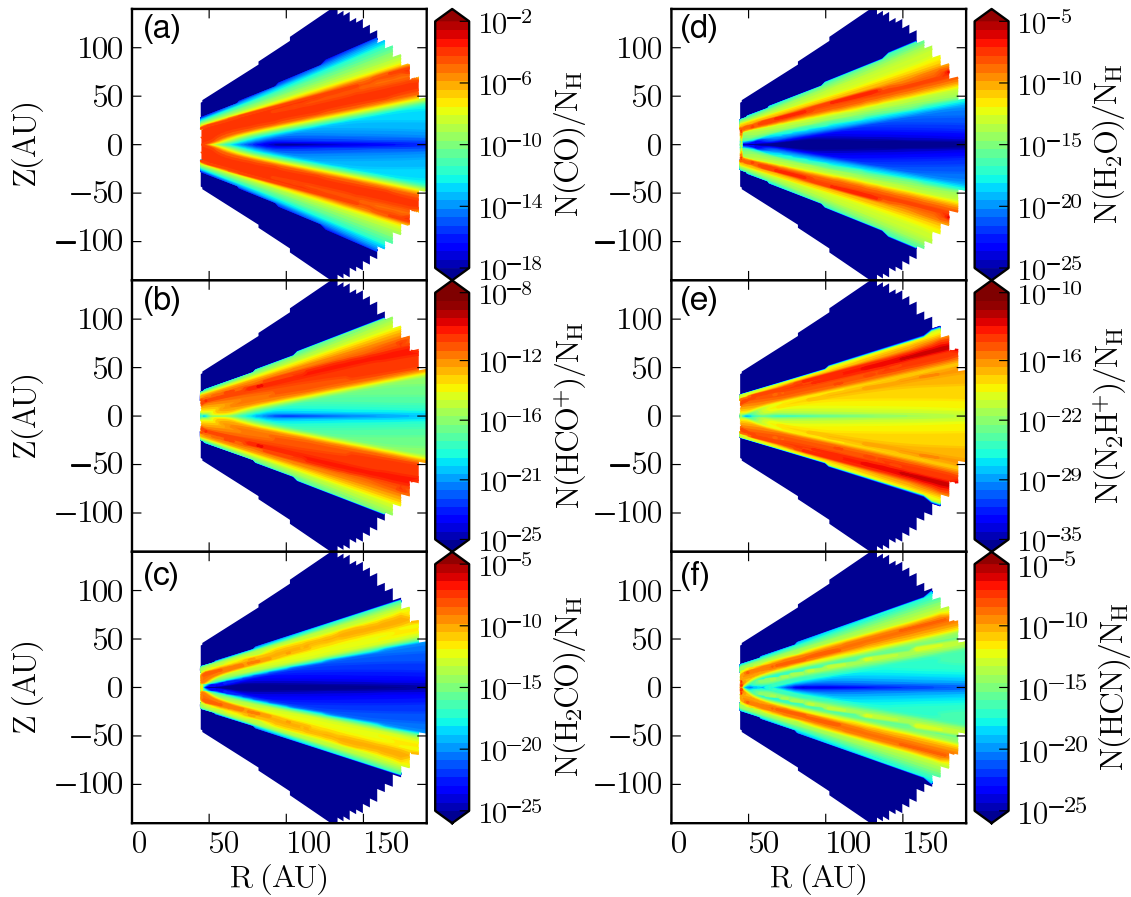}
\caption{Chemical model results, plotted as abundance relative to the total number of hydrogen atoms.  Shown are common species of astrophysical interest:  (a) CO, (b) HCO+, (c) H$_2$CO, (d) H$_2$O, (e) N$_2$H$^+$, (f) HCN.  \label{fig4}}
\end{figure*}

The resultant chemical abundances (relative to $n_{\rm H}=n_{\rm HI}+2n_{\rm H_{2}}$) for six observed gas-phase species are shown in Fig. \ref{fig4}.  Molecules such as CO, H$_{2}$CO, and N$_{2}$H$^{+}$ show an enhanced gas-phase abundance at the wall.  For comparison, in an untruncated disk model, all neutral species plotted in Fig. \ref{fig4} would be otherwise frozen onto grains at the midplane at R$_{\rm wall} = 45$ AU. The two ions shown, N$_{2}$H$^{+}$ and HCO$^{+}$, would also not be present in an untruncated disk, as their chemical precursors, N$_{2}$ and CO, would be frozen out, inhibiting the formation of these two species.  Consequently the gas-phase enhancement at the truncation radius shown for all species plotted here, besides H$_2$O, would not exist if it were not for the large inner gap. 

Water, with a freeze-out temperature of $\sim100$ K, is frozen onto grains at the $\sim50$ K transition region.  One could however envision a disk with a wall closer to the central star and thus warmer, such that water would sublimate from grains and be observable. Thus the specific species present are not necessarily the most important result, but that the star can efficiently heat the wall, warming it above the sublimation temperature of a variety of species, potentially allowing us to observationally probe disk physics as well as evolutionary state.  
Therefore the existence of a cleared inner gap should produce unique chemical features in the outer disk not present in full classical disks.

\subsection{Observables} \label{sec:obs}
If such an enhancement is present it is of interest to determine observability.  While previous observations reveal a diverse chemistry \citep[e.g. ][]{dutrey1997,oberg2010}, the sensitivity and resolving power of ALMA is required to fully understand the detailed structure of these systems.  

\begin{figure*}
\epsscale{1}
\plotone{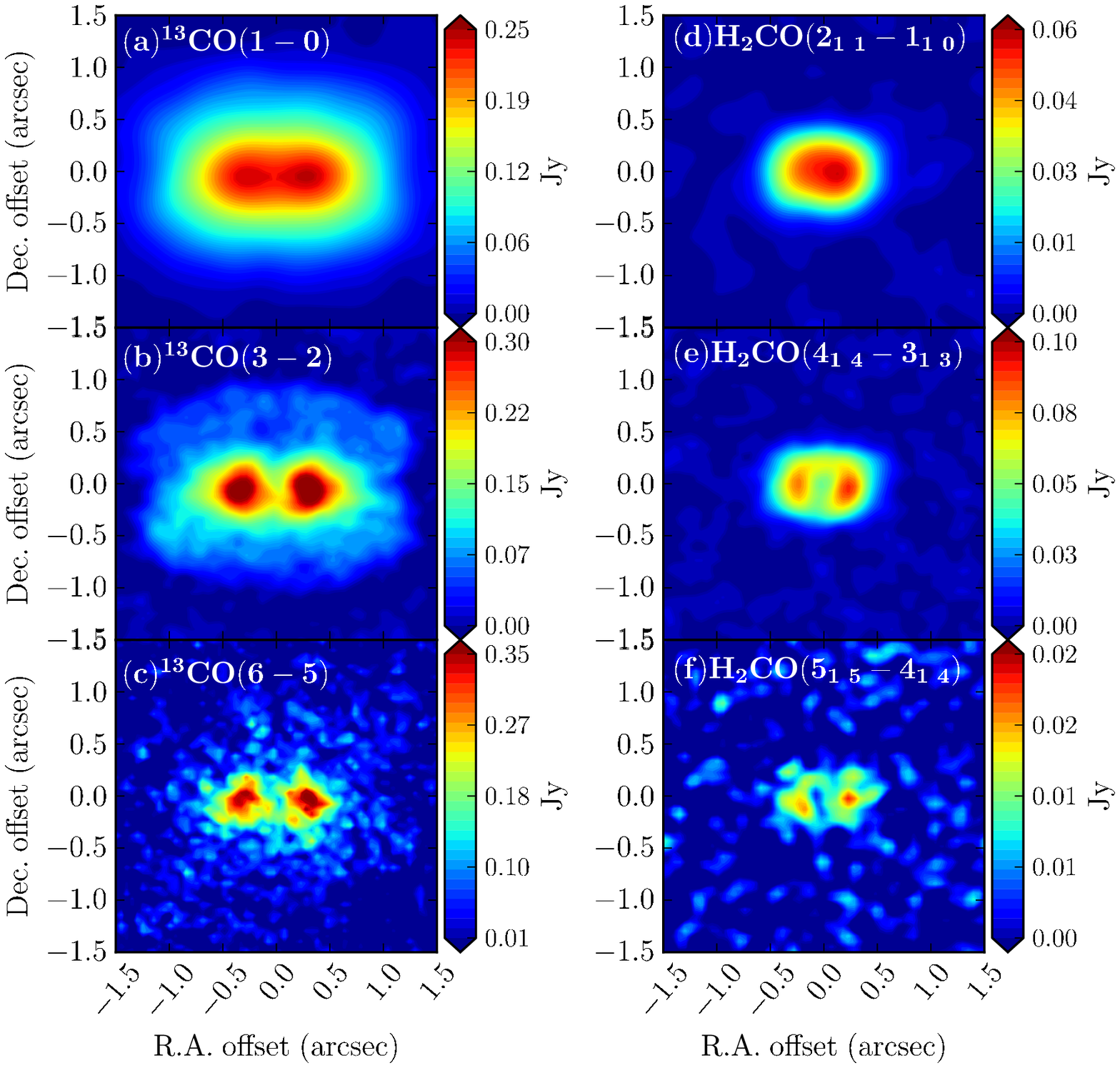}
\caption{Simulated ALMA observations (computed with one hour integration time) of $^{13}$CO and H$_2$CO for a disk at 140 pc, at different rotational transitions with increasing J.  This demonstrates the increasing contrast between the inner transition region and outer disk emission for higher J, along with ALMA's future capabilities at these frequencies.  \label{fig5}}
\end{figure*}

We adopt a disk inclination of $60^\circ$ and calculate the resulting emission for rotational transitions of $^{13}$CO and o-H$_2$CO as would be seen at a distance of $140$ pc using the non-LTE line radiation transfer code, LIME \citep{brinch2010}.  These species have been chosen as both are commonly observed towards disks around low mass T Tauri stars \citep[e.g. ][]{oberg2010,oberg2011}. The densities reached at the frontally illuminated midplane preferentially excite high-J rotational transitions, and therefore transition disks should uniquely exhibit high-J bright molecular rings at the wall when resolved by sub-mm observations with ALMA.  Using this calculated line emission, we then use the \textit{SIMDATA} package in CASA to calculate the ALMA visibilities and reconstruct the image from the UV coverage of the full ALMA array.  Fig. \ref{fig5} shows the simulated observations for these lines as seen by ALMA for an antenna configuration with $\sim0.1''$ resolution at 672 GHz, chosen to provide adequate resolution and sensitivity assuming typical thermal noise.  

\section{Further Considerations}

In this work we propose the presence of gas-phase molecules at the inner edge of transition disks.  This effect could however be erased if either the dust at the inner edge is not sufficiently heated or if the wall becomes permissive to molecule destroying UV radiation. In the following we discuss potential caveats of the model and how these would alter the results presented here.  

\subsection{Accretion Heating} \label{sec:accre}
The original model was computed assuming passive heating by the central star and heating due to disk accretion was treated as negligible.  Accretion heating, however, predominantly increases the midplane temperature \citep{dalessio1998}, and, if significant, will increase the thermal desorption of molecules from grains, enhancing the predicted effect.  Details of the dynamical transport of gas and dust and their effects on the chemistry, however, are beyond the scope of this Letter.

\subsection{Dust Settling and Grain Coagulation} \label{sec:dustgr}
For this model we have assumed an unsettled outer disk with grains of ISM abundance.  In our analysis we have also explored a model that includes mixed grain growth of dust particles of up to 1 mm in size, without vertical settling.  This lowers the UV opacity (cm$^2$ g$^{-1}$) by $\sim$ 2 orders of magnitude, which is a direct consequence of the fact that large grains contribute significantly to the mass but little to the UV opacity.  The high model densities at the inner edge ($\rho_{\rm dust}\sim10^{-14}$g cm$^{3}$), however, are still of sufficient magnitude such that the wall remains optically thick to UV radiation for a model with mixed grain growth alone.

Strong settling or removal of small grains can also make the disk more permeable to the UV radiation.  We can ask how much dust would need to be removed/settled to allow the UV photons to penetrate and photodissociate the enhanced abundance of molecules present at the wall. To approximate this, the unsettled wall density is $\rho_{\rm dust} \sim 10^{-14}$ g cm$^{3}$ and the UV dust opacity is $\sim 5 \times 10^{4}$ cm$^2$ g$^{-1}$ at 1000 \AA\ \citep{WD2001}.  Based on the thermal model, the thickness of the ``warm'' inner edge is $\sim$ 5 AU at R = 45 AU.  For the UV to penetrate the inner 5 AU, the disk would need contain $< 0.1\%$ of the original amount of dust present to erase the effect seen here.  We note however that settling will also increase the depth of the heating, which will in principle thicken the extent of the wall further.

\subsection{Dust in the Inner Gap} \label{sec:innerdust}
An underlying feature of the disk model is the presence of a large inner ``void" which allows the outer disk to be heated directly by the star.  One possible explanation for such a void is grain growth into rocky planetesimals \citep[e.g.][]{skrutskie1990}.  Thus one could postulate the existence of a small amount of undetected dust inside the gap.  Indeed some models infer the presence of some moderate mass of silicates in the inner disk \citep[e.g.][]{calvet2002,calvet2005,espaillat2007}.  This material has the ability to shadow the outer disk from being directly heated by the star, thus inhibiting the presence of the warm molecular interface presented here.  

First, one can ask how much dust would be required to cause the gap to become opaque ($\tau \sim10$) to optical radiation peaking at $\lambda=0.7$ $\mu$m as is the case for a $T=4000$ K star.  If we assume that the inner disk follows the outer disk profile (Section~\ref{sec:mod}), a total mass in dust of M$_{\rm dust}=10^{-3}$ M$_{\rm Moon}$ inside 45 AU would be sufficient for the midplane to become optically thick to the optical heating radiation. While this is an extremely small amount of dust, one must also consider the scale height of the material: $h\sim3.3$ AU at R = 45 AU.  Therefore, this cross-section is very small, and would not strongly shadow the outer disk and would not significantly alter the heating and UV irradiation except at the very central midplane region.  Furthermore, because the original model from \citet{sauter2009} included scattered light images, the presence of small grains at or above the disk scale height would significantly contribute to the amount of scattered light and not match observations for the model presented here.

\subsection{Gas in the Inner Gap}\label{sec:innergas}
By assuming gas and dust are co-spatial we intrinsically assume the gap is empty of gas.  Gas within the gap would not alter the original optical heating calculation of the outer disk or the propagation of UV continuum photons, both of which have dust-dominated opacities.  However, the presence of gas would strongly affect the propagation of \La\ as gas within the gap would likely be predominantly atomic.  The presence of hydrogen within the gap would cause a net reduction in the \La\ flux reaching the outer disk.  This would either not change or marginally enhance the abundance of gas phase molecules at the truncation radius since the amount of photo-dissociating radiation is reduced.  

\section{Conclusions} \label{sec:conclu}
We have investigated the chemistry of a transition disk employing the model structure of \citet{sauter2009} and have shown that a disk with a 45 AU inner void depleted of dust and gas should have unique chemical properties.  At the truncation radius, the midplane is both heated by optical photons and irradiated by the stellar UV and X-rays.  The net effect is that the previously hidden midplane is revealed as a region rich in gas-phase molecules, that should be readily observable by ALMA.  For comparison, in a full untruncated disk model, the midplane is normally cold and shielded from the central star, resulting in substantial freeze-out of gas onto grains, preventing direct observation.  While this analysis focuses on a specific disk model, these first results provide insight into our ability to directly probe the unique chemistry of transition disks.  The rich chemistry at the transition region will allow us to probe the physical conditions, such as density, temperature and kinematics. It is important to note that this model provides one possibility, which can be readily tested with ALMA.  We will be exploring a suite of disk models with a variety of geometric frameworks both in dust and gas content in future work, to further explore the full parameter space of transition disks, with the aim of pioneering future observational campaigns with ALMA to explore these exciting objects.

We thank the anonymous referee for their helpful comments.  This work was partially supported by NSF grant AST-1008800. J.S. acknowledges support by the DFG through the research group 759 ``The formation of planets: The critical first growth phase.''


\begin{thebibliography}{42}
\expandafter\ifx\csname natexlab\endcsname\relax\def\natexlab#1{#1}\fi

\bibitem[{{Aikawa} \& {Herbst}(1999)}]{aikawa1999}
{Aikawa}, Y. \& {Herbst}, E. 1999, \aap, 351, 233

\bibitem[{{Aikawa} \& {Herbst}(2001)}]{aikawa2001}
---. 2001, \aap, 371, 1107

\bibitem[{{Andrews} {et~al.}(2011){Andrews}, {Wilner}, {Espaillat}, {Hughes},
  {Dullemond}, {McClure}, {Qi}, \& {Brown}}]{andrews2011}
{Andrews}, S.~M., {Wilner}, D.~J., {Espaillat}, C., {Hughes}, A.~M.,
  {Dullemond}, C.~P., {McClure}, M.~K., {Qi}, C., \& {Brown}, J.~M. 2011, \apj,
  732, 42

\bibitem[{{Andrews} {et~al.}(2009){Andrews}, {Wilner}, {Hughes}, {Qi}, \&
  {Dullemond}}]{andrews2009}
{Andrews}, S.~M., {Wilner}, D.~J., {Hughes}, A.~M., {Qi}, C., \& {Dullemond},
  C.~P. 2009, \apj, 700, 1502

\bibitem[{{Bergin} {et~al.}(2003){Bergin}, {Calvet}, {D'Alessio}, \&
  {Herczeg}}]{bergin2003}
{Bergin}, E., {Calvet}, N., {D'Alessio}, P., \& {Herczeg}, G.~J. 2003, \apjl,
  591, L159

\bibitem[{{Bethell} \& {Bergin}(2011)}]{bethell2011}
{Bethell}, T.~J. \& {Bergin}, E.~A. 2011, \apj, 739, 78

\bibitem[{{Brinch} \& {Hogerheijde}(2010)}]{brinch2010}
{Brinch}, C. \& {Hogerheijde}, M.~R. 2010, \aap, 523, A25+

\bibitem[{Brown {et~al.}(2008)Brown, Blake, Qi, Dullemond, \&
  Wilner}]{Brown2008}
Brown, J.~M., Blake, G.~A., Qi, C., Dullemond, C.~P., \& Wilner, D.~J. 2008,
  The Astrophysical Journal Letters, 675, L109

\bibitem[{{Bryden} {et~al.}(1999){Bryden}, {Chen}, {Lin}, {Nelson}, \&
  {Papaloizou}}]{bryden1999}
{Bryden}, G., {Chen}, X., {Lin}, D.~N.~C., {Nelson}, R.~P., \& {Papaloizou},
  J.~C.~B. 1999, \apj, 514, 344

\bibitem[{{Calvet} {et~al.}(2002){Calvet}, {D'Alessio}, {Hartmann}, {Wilner},
  {Walsh}, \& {Sitko}}]{calvet2002}
{Calvet}, N., {D'Alessio}, P., {Hartmann}, L., {Wilner}, D., {Walsh}, A., \&
  {Sitko}, M. 2002, \apj, 568, 1008

\bibitem[{{Calvet} {et~al.}(2005){Calvet}, {D'Alessio}, {Watson},
  {Franco-Hern{\'a}ndez}, {Furlan}, {Green}, {Sutter}, {Forrest}, {Hartmann},
  {Uchida}, {Keller}, {Sargent}, {Najita}, {Herter}, {Barry}, \&
  {Hall}}]{calvet2005}
{Calvet}, N., {D'Alessio}, P., {Watson}, D.~M., {Franco-Hern{\'a}ndez}, R.,
  {Furlan}, E., {Green}, J., {Sutter}, P.~M., {Forrest}, W.~J., {Hartmann}, L.,
  {Uchida}, K.~I., {Keller}, L.~D., {Sargent}, B., {Najita}, J., {Herter},
  T.~L., {Barry}, D.~J., \& {Hall}, P. 2005, \apjl, 630, L185

\bibitem[{{Cieza} {et~al.}(2010){Cieza}, {Schreiber}, {Romero}, {Mora},
  {Merin}, {Swift}, {Orellana}, {Williams}, {Harvey}, \& {Evans}}]{cieza2010}
{Cieza}, L.~A., {Schreiber}, M.~R., {Romero}, G.~A., {Mora}, M.~D., {Merin},
  B., {Swift}, J.~J., {Orellana}, M., {Williams}, J.~P., {Harvey}, P.~M., \&
  {Evans}, N.~J. 2010, \apj, 712, 925

\bibitem[{{Cossins} {et~al.}(2010){Cossins}, {Lodato}, \&
  {Testi}}]{cossins2010}
{Cossins}, P., {Lodato}, G., \& {Testi}, L. 2010, \mnras, 407, 181

\bibitem[{{D'Alessio} {et~al.}(1998){D'Alessio}, {Canto}, {Calvet}, \&
  {Lizano}}]{dalessio1998}
{D'Alessio}, P., {Canto}, J., {Calvet}, N., \& {Lizano}, S. 1998, \apj, 500,
  411

\bibitem[{{Dullemond} {et~al.}(2001){Dullemond}, {Dominik}, \&
  {Natta}}]{dullemond2001}
{Dullemond}, C.~P., {Dominik}, C., \& {Natta}, A. 2001, \apj, 560, 957

\bibitem[{{Dutrey} {et~al.}(1997){Dutrey}, {Guilloteau}, \&
  {Guelin}}]{dutrey1997}
{Dutrey}, A., {Guilloteau}, S., \& {Guelin}, M. 1997, \aap, 317, L55

\bibitem[{{Dutrey} {et~al.}(1994){Dutrey}, {Guilloteau}, \&
  {Simon}}]{dutrey1994}
{Dutrey}, A., {Guilloteau}, S., \& {Simon}, M. 1994, \aap, 286, 149

\bibitem[{{Espaillat} {et~al.}(2007){Espaillat}, {Calvet}, {D'Alessio},
  {Bergin}, {Hartmann}, {Watson}, {Furlan}, {Najita}, {Forrest}, {McClure},
  {Sargent}, {Bohac}, \& {Harrold}}]{espaillat2007}
{Espaillat}, C., {Calvet}, N., {D'Alessio}, P., {Bergin}, E., {Hartmann}, L.,
  {Watson}, D., {Furlan}, E., {Najita}, J., {Forrest}, W., {McClure}, M.,
  {Sargent}, B., {Bohac}, C., \& {Harrold}, S.~T. 2007, \apjl, 664, L111

\bibitem[{{Espaillat} {et~al.}(2010){Espaillat}, {D'Alessio}, {Hern{\'a}ndez},
  {Nagel}, {Luhman}, {Watson}, {Calvet}, {Muzerolle}, \&
  {McClure}}]{espaillat2010}
{Espaillat}, C., {D'Alessio}, P., {Hern{\'a}ndez}, J., {Nagel}, E., {Luhman},
  K.~L., {Watson}, D.~M., {Calvet}, N., {Muzerolle}, J., \& {McClure}, M. 2010,
  \apj, 717, 441

\bibitem[{{Fogel} {et~al.}(2011){Fogel}, {Bethell}, {Bergin}, {Calvet}, \&
  {Semenov}}]{fogel2011}
{Fogel}, J.~K.~J., {Bethell}, T.~J., {Bergin}, E.~A., {Calvet}, N., \&
  {Semenov}, D. 2011, \apj, 726, 29

\bibitem[{{Glassgold} {et~al.}(1997){Glassgold}, {Najita}, \&
  {Igea}}]{glassgold1997}
{Glassgold}, A.~E., {Najita}, J., \& {Igea}, J. 1997, \apj, 480, 344

\bibitem[{{Glassgold} {et~al.}(2004){Glassgold}, {Najita}, \&
  {Igea}}]{glassgold2004}
---. 2004, \apj, 615, 972

\bibitem[{{G{\"u}nther} {et~al.}(2007){G{\"u}nther}, {Schmitt}, {Robrade}, \&
  {Liefke}}]{gunther2007}
{G{\"u}nther}, H.~M., {Schmitt}, J.~H.~M.~M., {Robrade}, J., \& {Liefke}, C.
  2007, \aap, 466, 1111

\bibitem[{{Herczeg} {et~al.}(2002){Herczeg}, {Linsky}, {Valenti},
  {Johns-Krull}, \& {Wood}}]{herczeg2002}
{Herczeg}, G.~J., {Linsky}, J.~L., {Valenti}, J.~A., {Johns-Krull}, C.~M., \&
  {Wood}, B.~E. 2002, \apj, 572, 310

\bibitem[{{Herczeg} {et~al.}(2004){Herczeg}, {Wood}, {Linsky}, {Valenti}, \&
  {Johns-Krull}}]{Herczeg2004}
{Herczeg}, G.~J., {Wood}, B.~E., {Linsky}, J.~L., {Valenti}, J.~A., \&
  {Johns-Krull}, C.~M. 2004, \apj, 607, 369

\bibitem[{{Hughes} {et~al.}(2009){Hughes}, {Andrews}, {Espaillat}, {Wilner},
  {Calvet}, {D'Alessio}, {Qi}, {Williams}, \& {Hogerheijde}}]{hughes2009}
{Hughes}, A.~M., {Andrews}, S.~M., {Espaillat}, C., {Wilner}, D.~J., {Calvet},
  N., {D'Alessio}, P., {Qi}, C., {Williams}, J.~P., \& {Hogerheijde}, M.~R.
  2009, \apj, 698, 131

\bibitem[{{Jonkheid} {et~al.}(2004){Jonkheid}, {Faas}, {van Zadelhoff}, \& {van
  Dishoeck}}]{jonkheid2004}
{Jonkheid}, B., {Faas}, F.~G.~A., {van Zadelhoff}, G.-J., \& {van Dishoeck},
  E.~F. 2004, \aap, 428, 511

\bibitem[{{{\"O}berg} {et~al.}(2010){{\"O}berg}, {Qi}, {Fogel}, {Bergin},
  {Andrews}, {Espaillat}, {van Kempen}, {Wilner}, \& {Pascucci}}]{oberg2010}
{{\"O}berg}, K.~I., {Qi}, C., {Fogel}, J.~K.~J., {Bergin}, E.~A., {Andrews},
  S.~M., {Espaillat}, C., {van Kempen}, T.~A., {Wilner}, D.~J., \& {Pascucci},
  I. 2010, \apj, 720, 480

\bibitem[{{{\"O}berg} {et~al.}(2011){{\"O}berg}, {Qi}, {Fogel}, {Bergin},
  {Andrews}, {Espaillat}, {Wilner}, {Pascucci}, \& {Kastner}}]{oberg2011}
{{\"O}berg}, K.~I., {Qi}, C., {Fogel}, J.~K.~J., {Bergin}, E.~A., {Andrews},
  S.~M., {Espaillat}, C., {Wilner}, D.~J., {Pascucci}, I., \& {Kastner}, J.~H.
  2011, \apj, 734, 98

\bibitem[{{Pi{\'e}tu} {et~al.}(2006){Pi{\'e}tu}, {Dutrey}, {Guilloteau},
  {Chapillon}, \& {Pety}}]{pietu2006}
{Pi{\'e}tu}, V., {Dutrey}, A., {Guilloteau}, S., {Chapillon}, E., \& {Pety}, J.
  2006, \aap, 460, L43

\bibitem[{{Rice} {et~al.}(2003){Rice}, {Wood}, {Armitage}, {Whitney}, \&
  {Bjorkman}}]{rice2003}
{Rice}, W.~K.~M., {Wood}, K., {Armitage}, P.~J., {Whitney}, B.~A., \&
  {Bjorkman}, J.~E. 2003, \mnras, 342, 79

\bibitem[{{Sauter} {et~al.}(2009){Sauter}, {Wolf}, {Launhardt}, {Padgett},
  {Stapelfeldt}, {Pinte}, {Duch{\^e}ne}, {M{\'e}nard}, {McCabe}, {Pontoppidan},
  {Dunham}, {Bourke}, \& {Chen}}]{sauter2009}
{Sauter}, J., {Wolf}, S., {Launhardt}, R., {Padgett}, D.~L., {Stapelfeldt},
  K.~R., {Pinte}, C., {Duch{\^e}ne}, G., {M{\'e}nard}, F., {McCabe}, C.,
  {Pontoppidan}, K., {Dunham}, M., {Bourke}, T.~L., \& {Chen}, J. 2009, \aap,
  505, 1167

\bibitem[{{Skrutskie} {et~al.}(1990){Skrutskie}, {Dutkevitch}, {Strom},
  {Edwards}, {Strom}, \& {Shure}}]{skrutskie1990}
{Skrutskie}, M.~F., {Dutkevitch}, D., {Strom}, S.~E., {Edwards}, S., {Strom},
  K.~M., \& {Shure}, M.~A. 1990, \aj, 99, 1187

\bibitem[{{Smith} {et~al.}(2004){Smith}, {Herbst}, \& {Chang}}]{smith2004}
{Smith}, I.~W.~M., {Herbst}, E., \& {Chang}, Q. 2004, \mnras, 350, 323

\bibitem[{{Strom} {et~al.}(1989){Strom}, {Strom}, {Edwards}, {Cabrit}, \&
  {Skrutskie}}]{strom1989}
{Strom}, K.~M., {Strom}, S.~E., {Edwards}, S., {Cabrit}, S., \& {Skrutskie},
  M.~F. 1989, \aj, 97, 1451

\bibitem[{{Umebayashi} \& {Nakano}(1981)}]{umebayashi1981}
{Umebayashi}, T. \& {Nakano}, T. 1981, \pasj, 33, 617

\bibitem[{{van Zadelhoff} {et~al.}(2003){van Zadelhoff}, {Aikawa},
  {Hogerheijde}, \& {van Dishoeck}}]{vanzadelhoff2003}
{van Zadelhoff}, G.-J., {Aikawa}, Y., {Hogerheijde}, M.~R., \& {van Dishoeck},
  E.~F. 2003, \aap, 397, 789

\bibitem[{{Weingartner} \& {Draine}(2001)}]{WD2001}
{Weingartner}, J.~C. \& {Draine}, B.~T. 2001, \apj, 548, 296

\bibitem[{{Willacy} \& {Langer}(2000)}]{willacyandlanger2000}
{Willacy}, K. \& {Langer}, W.~D. 2000, \apj, 544, 903

\bibitem[{{Wolf} \& {D'Angelo}(2005)}]{wolf2005}
{Wolf}, S. \& {D'Angelo}, G. 2005, \apj, 619, 1114

\bibitem[{{Wolf} {et~al.}(2002){Wolf}, {Gueth}, {Henning}, \&
  {Kley}}]{wolf2002}
{Wolf}, S., {Gueth}, F., {Henning}, T., \& {Kley}, W. 2002, \apjl, 566, L97

\bibitem[{{Wolf} {et~al.}(1999){Wolf}, {Henning}, \& {Stecklum}}]{wolf1999}
{Wolf}, S., {Henning}, T., \& {Stecklum}, B. 1999, \aap, 349, 839

\end{thebibliography}
\end{document}